# Adsorption-Induced Deformation of a Nanoporous Material: Influence of the Fluid-Adsorbent Interaction and Surface Freezing on the Pore-Load Modulus Measurement.

J. Puibasset[1*]

[1] Interfaces, Confinement, Matériaux et Nanostructures (ICMN), UMR 7374, CNRS et Université d'Orléans, 1b rue de la Férollerie, CS 40059, 45071 Orléans cedex 02, France

**Corresponding author:** E-mail: puibasset@cnrs-orleans.fr; Telephone: 33 2 38 25 58 03

# Abstract:

Liquid adsorption in nanoporous materials induces their deformation due to strong capillary forces. The linear relationship between the liquid pressure and the solid strain (pore-load modulus) provides an experimental technique to determine the mechanical properties of nanosized solids. Puzzling experimental results have often been reported, leading to a severe reconsideration of the mechanical properties of the thin walls, the introduction of surface stresses, and the suggestion of a mutual influence of fluid adsorption and matrix deformation. This work presents a molecular simulation examination of the fundamentals of the pore-load measurement technique. The pore-load protocol is reproduced as in experiments by measuring the solid deformation in presence of the liquid ("numerical experiment"), and the result is compared to the expected mechanical response of the solid. Focusing on a single nanoplatelet mimicking silicon stiffness, we show that the pore-load protocol is valid as long as the liquid in the pores remains liquid. However, when an ordered layer can form at the solid surface, it significantly affects the pore-load measurement. It is shown that this may happen above the freezing point even for moderately strong fluid-solid interactions. This observation could help for the interpretation of experimental data, in particular in porous silicon, where the expected presence of atomically smooth surfaces could favor the formation of highly ordered fluid layers.

# 1. Introduction

Recently, many theoretical and experimental results have been reported on the mechanical properties of nanostructures, in particular the elastic modulus of nanowires, nanoplates and nanobeams due to their widely proposed applications in nanoelectromechanical systems (NEMS). The knowledge of their mechanical properties is of considerable interest and has attracted many studies, which have revealed a strong size-dependence while approaching the nanometer scale. A straightforward procedure to understand the mechanical properties of small systems is probably to perform direct measurements at the nanoscale in simple geometries. A good example of model system to study size and surface effects is probably the case of silicon nanocantilever.[1-3] The size dependence of the mechanical properties has been evidenced, and has generally been attributed to surface stresses induced by the free surfaces.[4-10]

The same surface stresses are expected to play an important role in the deformation of nanoporous materials which exhibit large surface-to-volume ratio.[11] Furthermore, these materials can adsorb fluids, inducing capillary forces strong enough to deform the substrate. This concerns not only soft materials like aerogels or biopolymers,[12-18] but also stiff materials.[19-25] This adsorption-induced deformation[26] is a fundamental issue resulting from the interplay between the solid surface and the adsorbed fluid.[27-29] Beyond its importance for sequestration, storage or oil recovery,[30] or the fact that it may play a role in the shape of the adsorption isotherms,[31-40] it is certainly a powerful tool to measure the elastic moduli of nanoporous systems for which direct measurements are not always possible,[16,39,41-47] despite recent attempts.[48] The pore-load measurement is done as follows: a nanoporous material is saturated with a liquid and brought below its saturated vapor pressure. The resulting capillary tension in the liquid (negative pressure $P_\text{L}$) is large enough to induce measurable deformations of the nanoporous material. For isotropic porous materials, in the limit of elastic deformation $\varepsilon$, the pore-load modulus is defined as $M = dP_\text{L}/d\varepsilon$. This quantity, introduced by Biot in 1941,[49] measures the (linear) relation between the liquid pressure variations inside the pores and the solid deformation. It is an effective elastic modulus that depends on the solid stiffness but also on its geometry.[21,50-54] In anisotropic solids, it also depends on the direction.[45,46] This quantity provides in general not enough information to deduce the intrinsic elastic constants of the solid. It is however possible if complementary measurements can be performed.[47]

Poromechanics is a consistent approach to model the mechanical coupling between fluids and solids, including adsorption-induced deformations.[18,49,55-57] However, nanoporous materials are complex systems: comparison between experiments and theory is not straightforward, and more sophisticated models, including simulations, are required to take into account the inhomogeneous and anisotropic nature of the material.[21,22,29,54,58-69] It is well-known that porous materials have smaller elastic moduli than the bulk, mainly due to their lower density; but the exact geometry of the porous material needs to be considered in order to quantitatively predict the mechanical properties of the system.[51-53] There is however generally a disagreement between theory and experiments, which can be large, despite the development of refined models taking into account the pore geometry and surface stress.[11,40,41,43,47,54,70,71] Several hypotheses have been invoked, including surface defects,[3] or a possible interplay between surface stress and adsorption that could explain some adsorption hysteresis features.[31-40] In that case, one expects a fluid-dependence of the pore-load modulus. Such observations have been reported in the literature several decades ago,[32] but are not confirmed by recent experiments.[47]

The following question thus arises: is there an effect of the fluid-wall interaction on the adsorption-induced deformation in the saturation regime? Considering the small (nanometric) scale involved, molecular simulations provide the natural framework to investigate this issue. The idea is to perform a self-consistent test of the adsorption-induced deformation of a nanosized system (a nanoplatelet), by way of a "numerical experiment" that mimics a real experiment. The observed deformation is then compared with the mechanical properties of the solid. The key point is that the geometry of the solid and the forces between the atoms being known, all elastic constants of the solid, including the surface stresses, can be determined. The solid deformation in presence of an adsorbed fluid can thus be determined in the framework of a continuous model and compared with the observed deformation. Any difference would be the signature of a phenomenon beyond the surface stress of the solid. Varying the solid thickness and the intensity of the fluid-solid interaction helps to determine the origin of the observed discrepancy.

The paper is organized as follows: section 2 is devoted to the presentation of the atomistic model, the molecular simulation methods, as well as the thermodynamic model of solid deformation and the expected value for the pore-load modulus. Section 3 focuses on the simulation results for the fluid and solid physical properties, as well as the pore-load numerical experiment. The effect of the solid thickness and the intensity of the fluid-solid interaction are then discussed. It is shown that a modification of the fluid structure at the solid surface could explain the simulation results, and we conclude on the relevance of the phenomenon for experimentalists.

## 2. Methods

### 2.1. Model

Let us consider a nanoporous material in contact with a vapor at pressure $P$ below its saturated vapor pressure $P_{\text{sat}}$. Figure 1 shows a typical adsorption-desorption isotherm. Thermodynamics tells us that the fluid-solid interactions favor the condensed liquid phase when approaching the saturated vapor pressure: this is the well-known capillary condensation that occurs in nanometric pores. The isotherm exhibits a hysteresis: upon reduction of the gas pressure, the system remains saturated with liquid until desorption pressure ($P_{\text{des}}$) is reached. Starting from the saturated system, it is possible to travel back and forth reversibly along the saturation plateau, without emptying the nanoporous material. The pore-load modulus is measured along this plateau.

Let us consider a portion of a pore wall surrounded by liquid on both faces (see inset in Figure 1). This will define our simulation box (see Figure 2). For simplicity, we will consider a wall of constant thickness. The x and y directions are chosen parallel to the solid. The corresponding dimensions ($L_x$ and $L_y$) are chosen large enough, together with periodic boundary conditions, to minimize finite size effects in these directions. The dimension $L_z$ perpendicular to the solid surface is also chosen large enough so that the liquid far from the surface is bulk-like. Periodic boundary conditions are also applied in that direction to avoid the introduction of an extra interface. As a consequence, the fluid in the simulation box is actually confined between two flat surfaces (equivalent to a slit pore), which are far enough not to introduce cross-talking (the fluid in the center of the slit pore is bulk-like). This system can thus be seen as a nanoplatelet immersed in a liquid.

The introduction of periodic boundary conditions parallel to the solid wall is equivalent to considering an infinite porous material without explicit boundaries. As a consequence, we do not treat explicitly the liquid-vapor interfaces (meniscus) which are localized at the periphery of the porous material (saturation). We also implicitly disregard their propagation from the boundary towards the core of the material upon desorption. As a consequence, in our simulation model, it will be possible to decrease the liquid pressure down to its bulk stability limit (spontaneous cavitation). This pressure is generally well below the emptying pressure of the system, which, in most situations, is driven by meniscus propagation or heterogeneous cavitation, and strongly depends on the porous solid chemistry and structure. Another consequence of the absence of boundaries is that we can increase the liquid pressure above its saturated vapor pressure, which is not possible experimentally. These extra capabilities offered by simulation will be used to enlarge the range of variation of the liquid pressure, to explore new physics or increase the accuracy.

## 2.2. Simulation Methods

We use Monte Carlo simulations in different statistical ensembles and simulation box configurations to determine the equilibrium properties of our system in two situations: the dry nanoporous material, and the same system saturated with liquid at a given chemical potential (or equivalently liquid pressure). In both cases, the system is supposed to be in thermal equilibrium with a reservoir at temperature T.

(i) In the first configuration, only the solid nanoplatelet is present (there is no fluid, see Figure 2a). The dimensions $L_x$ and $L_y$ are allowed to fluctuate, while an external pressure is imposed in the corresponding directions. The dimension $L_z$ is kept constant. This corresponds to the isostress ensemble in directions x and y. This box configuration is used to determine the elastics constants of the solid, as follows. A given pressure is imposed along the direction x, while zero pressure is imposed in the directions y and z (along y, the box is left free to fluctuate, and the pressure on the solid in direction z is zero by construction). The measured deformations along directions x, y and z allow to calculate the Young modulus along x, and the Poisson's ratio $\nu_{xy}$ and $\nu_{xz}$. The in-plane modulus B can also be obtained by imposing the same pressure along the x and y directions. The other parameters of the compliance matrix will not be used.

(ii) In the second box configuration, the liquid is introduced, in equilibrium with a reservoir at chemical potential $\mu$ (see Figure 2b). It corresponds to the semi-grand canonical ensemble. The chemical potential of the reservoir is given by the gas pressure above the nanoporous material. In this box configuration, the external pressure in directions x and y is set equal to zero: the liquid pressure $P_L$ and the solid stress thus have to compensate on the simulation box walls, which induces a deformation of the solid. This mimics a pore-load modulus measurement where the nanoporous solid is left free to deform upon adsorption. The detailed analysis is given in section 2.4.

## 2.3. Interatomic Potentials

*2.3.1. The Solid.* The atomic structure of the nanoplate is chosen to be fcc. The interatomic solid-solid (s-s) potential is chosen to be the Lennard-Jones (12, 6)

$$V_{LJ}(r) = 4\varepsilon_{s\text{-}s}\left[\left(\frac{\sigma_{s\text{-}s}}{r}\right)^{12} - \left(\frac{\sigma_{s\text{-}s}}{r}\right)^{6}\right]. \tag{1}$$

with parameters $\varepsilon_{s\text{-}s}$= 73.2 kJ/mol and $\sigma_{s\text{-}s}$= 0.3518 nm. The interactions are cut at $4\sigma_{s\text{-}s}$ (see Table 1). This simple potential is able to reproduce the elastic behavior of a solid, including surface stress thanks to the long range Lennard-Jones interactions. More specifically, the numerical values of the parameters have been chosen to mimic the mechanical properties of silicon,[72] a material that has been used several times to study adsorption-induced deformation.[2,41,46,47,54] Note that silicon is not an fcc solid, and that better potentials exist to reproduce its physical properties.[73] However, an accurate description of the material is out of the scope of this study. The bulk properties have been determined in a cubic simulation box containing 6 unit cells (3.26 nm) in each direction with periodic boundary conditions (the crystallographic axes were parallel to the simulation box). The simulations have been done in the isostress ensemble where an external pressure can be applied independently in the three directions. The symmetry being cubic, and omitting the shear modulus, not used in this work, the mechanical properties of the bulk fcc solid are characterized by two parameters that have been calculated at 300K: the Young modulus E = 165 GPa, and the Poisson coefficient $\nu$ = 0.36 (see Table 2).

*2.3.2. The Fluid.* Since we focus on non-specific effects, we use again the Lennard-Jones (12,6) potential (eq 1) to model the fluid-fluid and fluid-solid interactions. Following Stoddard *et al.*,[74] the potential is truncated at $3\sigma_{x-x}$ and a quadratic term is added so that both the potential and force are continuous. The parameters are chosen to mimic n-heptane, a non-polar fluid which has been previously used for experiments.[46,75] The parameters have been determined based on the work of Watanabe *et al.*[76] The fluid-fluid parameters are $\sigma_{f-f}$ = 0.6 nm and $\varepsilon_{f-f}/k$ = 505 K, where k is Boltzmann's constant, or $\varepsilon_{f-f}$ = 4.2 kJ/mol. The fluid-solid parameters are taken equal to the fluid-fluid parameters $\varepsilon_{f-s} = \varepsilon_{f-f}$ and $\sigma_{f-s} = \sigma_{f-f}$. Note that $\varepsilon_{f-s}$ will be varied in section 3.4.1 to evidence fluid-solid interaction effects. The calculations cannot be done at 300 K because the Lennard-Jones fluid is solid at that temperature. The temperature is thus chosen to be 353 K, above the melting point of the Lennard-Jones model.

## 2.4. Pore-load Modulus: Thermodynamic Model

The liquid inside a pore may either be under positive or negative pressure depending on the chemical potential value. Without external forces, the general mechanical equilibrium equations impose that the stress along the pore is constant. The system being free, the integrated stress into the liquid compensates that in the solid (the gas pressure outside the pore is negligible, see below). As a consequence, the solid will deform, and the pore-load modulus measures the (linear) relation between the liquid pressure variations inside the pores and the solid deformation, to be determined now in the framework of the elastic theory.

Let us consider the system depicted in Figure 2b. We denote $P_L$ the pressure of the liquid and $\gamma$ the liquid-solid surface tension. At equilibrium, the external forces applying into the solid being zero, the internal stress in the three directions $\sigma_{xx}$, $\sigma_{yy}$ and $\sigma_{zz}$ are constant. Considering the symmetry of the nanoplate, one has $\sigma_{xx} = \sigma_{yy} \neq \sigma_{zz}$. Along the z axis, the stress into the solid equals that in the liquid, and in directions x and y, the overall force on each simulation box wall is zero since the external pressure is set to zero (free walls). As a consequence, the stress components into the solid are given by:

$$\sigma_{xx} = \sigma_{yy} = \frac{(HP_L - 2\gamma)}{h} \qquad 2$$

$$\sigma_{zz} = -P_L \qquad 3$$

where $h$ and $H$ are the solid and liquid thicknesses (defined in Figure 2). The deformation along the x and y directions ($\varepsilon_{xx} = \varepsilon_{yy}$) are then given by:

$$\varepsilon_{xx} = \varepsilon_{yy} = \frac{(1-\nu_{yx})\sigma_{xx}}{E_x} - \frac{\nu_{zx}\sigma_{zz}}{E_z} \qquad 4$$

$$= \left[\frac{(1-\nu_{yx})H}{hE_x} + \frac{\nu_{zx}}{E_z}\right]P_L - 2\gamma\frac{(1-\nu_{yx})}{hE_x} \qquad 5$$

where $E_x$, $E_z$, $\nu_{yx}$, and $\nu_{zx}$ are the elastic constants of the nanoplate. As can be seen, the fluid pressure acts on the solid deformation through two mechanisms materialized by the arrows in Figure 2b. (i) direct: a positive fluid pressure along z induces a nanoplatelet compression along z, and thus a dilatation along x and y given by the Poisson ratio $\nu_{zx}$. (ii) indirect: a positive pressure tends to dilate the simulation box in the x and y directions and thus the nanoplatelet. The surface tension is given by the excess free energy of the interface. In principle, it depends on the chemical potential $\mu$ of the fluid. However, its variations with $\mu$, given by that of the excess adsorbed liquid at the interface, are expected to be small. This point will be quantitatively discussed below (section 3.4.2). The solid deformation is thus expected to be essentially linear with the liquid pressure, and the pore-load modulus is given by:

$$M = \frac{dP_L}{d\varepsilon_{xx}} = \left[\frac{(1-\nu_{yx})H}{hE_x} + \frac{\nu_{zx}}{E_z}\right]^{-1} \qquad 6$$

As expected, the pore-load modulus depends on the geometry and the elastic constants of the solid. These quantities will be calculated independently (section 3.2).

## 3. Results and Discussion

### 3.1. Thermodynamic Properties of the Fluid

The determination of the pore-load modulus requires the knowledge of the liquid pressure $P_L$ far from the walls, *e.g.* in the center of the pores. It is assumed to be equal to that of the bulk liquid in equilibrium with the vapor at pressure $P$ or chemical potential $\mu$, given approximately by:

$$P_L \cong \frac{\mu - \mu_{\text{sat}}}{V_M} \cong \frac{RT}{V_M} \ln\left(\frac{P}{P_{\text{sat}}}\right) \qquad 7$$

where $R$ is the ideal gas constant and $V_M$ the molar volume of the liquid. This relation is frequently used, but molecular simulations can actually provide the accurate relation between $P_L$ and the activity $z = e^{\beta\mu}/\Lambda^3$, where $\Lambda$ is the de Broglie thermal wavelength and $\beta = 1/kT$. The results are given in Table 3 and Figure 3, as a function of ln($z$), which is approximately equal to $\ln(P)$ to within a constant. These results have been obtained by Grand Canonical Monte Carlo simulations and standard thermodynamic methods.[77,78] As can be seen, the liquid branch is not perfectly linear due to the small compressibility of the liquid, which justifies the exact calculation instead of using eq 7. Note that the range of pressures reached in the liquid is wide, from -122 to +202 bar. Below the saturated vapor point $P < P_{\text{sat}}$, $P_L$ is essentially negative, and decreases down to the stability limit of the liquid (bulk cavitation) which occurs at -122 bar. This region of negative pressures corresponds approximately to that used in experiments for pore-load measurements, except for the fact that in most cases the pore emptying occurs well above the bulk cavitation. Simulations will also be performed above $P_{\text{sat}}$, up to $P_L = 202$ bar: this will improve the statistics.

## 3.2. Mechanical Properties of the Nanoplatelet

The properties of the nanoplatelet are evaluated in a simulation box of initial size $L_x = L_y = 7$ unit cells in the x and y directions (3.805 nm); a gap is introduced along the z direction so as to create two opposite surfaces (Figure 2a). The thickness $h$ of the nanoplatelet is 6 unit cells ($h = 3.26$ nm), and the dimension $L_z = 10$ nm. The distance between the walls is thus $H = 6.74$ nm. The wall thickness is chosen small (3.26 nm) compared to typical nanoporous silicon walls (5-6 nm), in order to emphasize surface stress effects. The gap is however typical of nanoporous materials and large enough to avoid cross-talk through periodic boundary conditions. The elastic properties of the platelet have been determined in the framework of the standard Monte Carlo simulations in the isostress ensemble, where only the dimensions parallel to the nanoplatelet are allowed to vary, while $L_z$ is fixed. The direct measurements of the elastic properties of the solid allow taking into account the surface stresses induced by the long range interactions and the finite thickness of the platelet. Equations 5 and 6 require the knowledge of only three independent quantities: $E_x$, $\nu_{yx}$ and $\frac{\nu_{zx}}{E_z}$. Using the symmetry of the compliance matrix ($\nu_{yx} = \nu_{xy}$ and $\frac{\nu_{zx}}{E_z} = \frac{\nu_{xz}}{E_x}$), one can determine these parameters by applying uniaxial stress along x ($\sigma_{xx} \neq 0, \sigma_{yy} = \sigma_{zz} = 0$) and measuring the average deformations $\varepsilon_{xx}$, $\varepsilon_{yy}$ and $\varepsilon_{zz}$ (the fluctuations are discussed below). The linear regime extends up to $\pm 300$ bar, and one obtains $E_x = \frac{\sigma_{xx}}{\varepsilon_{xx}}$ = 162 GPa, $\nu_{xy} = -\frac{\varepsilon_{yy}}{\varepsilon_{xx}}$ = 0.457, and $\frac{E_x}{\nu_{xz}} = -\frac{\sigma_{xx}}{\varepsilon_{zz}}$ = 531 GPa (Table 2). As can be seen, the Young modulus is slightly lower than its bulk value, while Poisson's ratio is more significantly affected.[4-9] Furthermore, one can deduce $\nu_{xz}$ = 0.305, reflecting the nanoplate anisotropy. These values allow to calculate the expected pore-load modulus (eq 6) for the nanoplatelet: $M = 113$ GPa.

Let us now discuss the fluctuations and uncertainties. Figure 4a displays the fluctuations of the free nanoplate size $L_x$ along a simulation run. The fluctuations follow a Gaussian law of width $8 \times 10^{-3}$ nm (Figure 4c): this is quite large compared to the nanoplate dimensions, as expected in nanometric systems. The corresponding fluctuations in stress are large (of order $3 \times 10^3$ bar), and are actually much larger than the typical stress applied to the solid. As a consequence, long simulation runs are required to reach the desired relative accuracy for the average strain of the solid (5%).

The situation may be improved when one can take into account a natural symmetry of the system. This is the case for the free nanoplatelet (xy symmetry). The method consists in imposing the geometrical constraint $L_x = L_y$. The fluctuating value of $L_x = L_y$ is displayed in Figure 4b and c. As can be seen, the average value of is not affected, but the system being now stiffer, the fluctuations are smaller. The same simulation length will thus provide more accurate results. This procedure will be used during the pore-load measurements (in presence of liquid), thanks to the natural symmetry of the system.

## 3.3. Adsorption Induced Deformations

Let us now consider the case where the nanoplatelet is immersed in a liquid characterized by the logarithm of its activity $\ln(z)$. As previously, we have coupled the x and y directions ($L_x = L_y$) to improve the accuracy. The histograms associated to $L_x$ fluctuations are fitted with Gaussian distributions. The results are given in Figure 5 and Table 3. As can be seen, the Gaussian distributions overlap: the fluctuations are of the order of $10^{-3}$ $L_x$, while the deformation is of order $10^{-4}$ $L_x$, typical of solid strain. An interesting feature regarding fluctuations is that they are essentially independent of $\ln(z)$ for the wet solid, but are slightly larger than for the dry solid. The presence of the liquid in close interaction with the solid affects the amplitude of the fluctuations.

In presence of liquid, the system may either contract or dilate, depending on the liquid activity z. For high activity (positive liquid pressure), the solid expands, while for low activity (negative liquid pressure) the solid shrinks. This is qualitatively expected from the observation that the fluid pressure acts directly on the solid (along z-axis) and through the simulation box (parallel to the nanoplatelet). At coexistence ($\ln(z)$ = -10.46), the liquid pressure is essentially zero (equal to the vapor pressure). The observed deformation in that case is essentially due to the surface tension term in eq 5. The observed deformation is very small: the fluid-solid surface tension is thus small compared to the solid stiffness. This is closely connected to the fact that we have chosen fluid-wall interactions equal to fluid-fluid interactions.

To determine the pore-load modulus, the simulation data are drawn as a function of the liquid pressure in Figure 6. As can be seen, the strain follows essentially a linear behavior, giving a pore-load modulus of 246 GPa. The lowest pressure point corresponds to the stability limit of the stretched liquid. At this point, the probability to form transient gas bubbles is not negligible, and strongly dependent on the presence of a solid wall. This could explain why this point slightly deviates from the linear behavior. It is emphasized that the range of pressure accessible experimentally is narrower. The lower limit is given by the pressure where the porous solid empties, generally above the bulk cavitation limit, and the upper limit is given by the saturated vapor pressure of the fluid (close to zero, see Figure 3). The simulations show that the observed linear behavior extends beyond these experimental limits, in particular above the saturation point.

The pore-load modulus deduced from the simulation results (246 GPa) deviates significantly from the expected value given by the phenomenological model (113 GPa). For visualization, the prediction of the model (eq 5) is given as a solid line in Figure 5, where we have omitted the constant surface tension term which is weak anyway. The "numerical experiment" thus gives a pore-load modulus that is larger than expected from the elastic constants of the solid. Conversely, if one deduces the bulk modulus of the solid from the pore-load measurement, a large overestimation is done.

## 3.4. Discussion

Disagreements between the elastic moduli determined from adsorption-induced deformation measurements and bulk values have already been reported in the literature. In most cases, the origin is attributed to the surface stress effects due to the small wall thickness in nanoporous materials. Since in our simulations the elastic constants have been determined for the nanoplatelet itself, these finite size effects cannot be invoked to explain the disagreement.

On the other hand, one can invoke the dependence of the solid (surface) stress with the presence of the adsorbed fluid. Two hypotheses are proposed to explain the results: a strong dependence of the fluid-substrate free energy with the chemical potential of the fluid, or a significant variation of the surface stress of the solid in presence of the fluid. The second argument has already been invoked to explain some features of the nitrogen adsorption hysteresis in porous silicon.[35-38] In order to test these hypotheses, we have studied the influence of the fluid-solid interaction intensity and the effect of the solid thickness.

*3.4.1. Influence of Fluid-Solid Interactions and Solid Thickness.* The most direct route to evidence an effect due to the interface is to vary the intensity of the fluid-solid interaction and/or the nanoplate thickness. We have first considered a reduction of the interaction parameter $\varepsilon_{f-s}$ by a factor two to ten ($\varepsilon_{f-s}/\varepsilon_{f-f}$ = 1.0, 0.5, 0.25, and 0.1). The corresponding pore-load modulus has been calculated and plotted in Figure 7 as a function of $\varepsilon_{f-s}/\varepsilon_{f-f}$. The expected modulus from the thermodynamic model (eq 6 and Table 2) is given as a horizontal line. As can be seen, the modulus given by the "numerical experiment" exhibits a dependence on the fluid-solid interaction. More specifically, two regimes can be determined: $\varepsilon_{f\text{-}s} \leq 0.25\, \varepsilon_{f\text{-}f}$ where the modulus is constant and equal to the expected value deduced from the mechanical properties of the nanoplatelet (113 GPa), and $\varepsilon_{f\text{-}s} \geq 0.5\, \varepsilon_{f\text{-}f}$ where the modulus is found to be significantly larger than expected (more than a factor two) with a dependence on $\varepsilon_{f\text{-}s}$. The extrapolation between $0.25\, \varepsilon_{f\text{-}f}$ and $0.5\, \varepsilon_{f\text{-}f}$ suggests a sharp transition between the two regimes.

Let us now focus on the effect of the nanoplatelet thickness. The idea is to determine whether the disagreement between the numerical experiments and theory is due to a volume or a surface contribution from the nanoplatelet. We proceed as follows. We perform simulations for two nanoplatelet thicknesses ($h$ = 1.63 nm and 3.26 nm). The results are then analyzed with eq 5 which contains the surface excess free energy $\gamma$. This term was previously discarded to obtain the pore-load modulus as given by eq 6. Let us now suppose conversely that the discrepancy between simulation results and theory originates purely from the surface term $\gamma$, and let us deduce its value from the simulation data for the two nanoplatelet thicknesses (see Figure 8). As can be seen, the points fall on the same curve within errors, proving that the discrepancy arises from a surface contribution. Figure 8 also shows that this surface term strongly depends on the chemical potential of the fluid. It is the variations of this surface term with the liquid pressure which contributes to the pore-load modulus. The physical origin of this surface contribution is discussed now.

### 3.4.2. Fluid-Solid Free Energy.
In order to check the thermodynamic consistency of the surface term previously deduced from the solid deformations, we evaluate the fluid-solid free energy (surface tension) through another route.[79-81] The Gibbs equation gives the variations of the excess (surface) free energy versus chemical potential variations:

$$d\gamma^{\text{Gibbs}} = -\Gamma^{\text{ex}} d\mu = -T\Gamma^{\text{ex}} d\ln(z) \qquad \qquad 8$$

where $\Gamma^{\text{ex}}$ is the excess adsorbed fluid relative to the liquid state, $T$ the temperature, $\mu$ the chemical potential and $z$ the activity of the adsorbed fluid. $\gamma^{\text{Gibbs}}$ actually depends on the position of the Gibbs surface, and eq 8 results from the standard choice corresponding to a zero adsorbed excess for the solid. Figure 9 gives the results for the excess adsorbed fluid: it is a positive value since the attractive fluid-solid interaction causes the fluid to be slightly denser close to the surface. The magnitude is however small due to the low compressibility of the liquid. Integration of eq 8 from the (arbitrary) reference point $\ln(z) = -10.485$ gives the corresponding $\gamma^{\text{Gibbs}}$ to within a constant (see Figure 9). Comparison with Figure 8 immediately shows strong disagreements. Quantitatively, the free energy variation found by the thermodynamic route is one order of magnitude smaller than the value required to explain the mechanical deformation of the nanoplatelet in presence of the fluid, and with the wrong sign. As a consequence the observed discrepancy cannot be explained in terms of fluid-wall excess free energy.

*3.4.3. Fluid Structure at the Interface.* It is known that the fluid structure at the interface with a solid may depart significantly from the bulk due to the strong interactions with the substrate.[82-86] The atomic structure of the fluid can be revealed by measuring density profiles along the simulation runs. The results are given in Figure 10. Visual inspection of the atomic configurations reveals that the fluid is highly structured at the interface, due to the fluid-solid interaction and the flatness of the interface. The local density profiles confirm this observation, and show that the fluid ordering may extend to two layers. The fluid structure has been determined for various interaction intensities, and exhibits significantly higher ordering for $\varepsilon_{f-s} \geq 0.5\ \varepsilon_{f-f}$: the peaks are sharp, in particular for the first layer, and the fluid density in the interlayers reaches zero, while for $\varepsilon_{f-s} \leq 0.25\ \varepsilon_{f-f}$, the peaks are rounded and the interlayers are partially filled. The snapshots of the first atomic layer of the fluid show long range ordering for $\varepsilon_{f-s} \geq 0.5\ \varepsilon_{f-f}$ that are absent for $\varepsilon_{f-s} \leq 0.25\ \varepsilon_{f-f}$. The fluid crystallizes at the interface due to the fluid-solid interaction and the flat surface that favors long range ordering. This phenomenon has been quantitatively described by Radhakrishnan et al.[85] who have proposed a global phase diagram. They have introduced a quantity $\alpha$ which measures the strength of attraction of the pore walls relative to the fluid-fluid interaction, $\alpha = \left(\rho_s \varepsilon_{f-s} \sigma_{f-s}^2 \Delta\right)/\varepsilon_{f-f}$ where $\rho_s$ is the solid density, $\Delta$ the distance between the atomic layers of the solid wall and $\varepsilon_{f-s}$ and $\sigma_{f-s}$ are the fluid-solid parameters. Numerical evaluation for our system gives $\alpha = 2.44 \varepsilon_{f-s}/\varepsilon_{f-f}$. The authors have shown that the first fluid layer in contact with the solid may "freeze" above the bulk freezing point for $\alpha$ roughly larger than 1, which corresponds to $\varepsilon_{f-s}/\varepsilon_{f-f}$ larger than 0.4. Our simulation results are in agreement with this result. It is proposed that this surface ordering could explain the unexpected deformation of the nanoplatelet, for the following reasons:

(i) The mechanical properties of the ordered (frozen) fluid at the interface are expected to be very different from that of the bulk fluid, in particular regarding the relation between the chemical potential and the mechanical pressure. We thus expect a modification of the mechanical response of the system when the first fluid layer crystallizes. This solid layer is expected to increase the effective stiffness of the nanoplatelet, as observed in the simulations.

(ii) The crystallization occurs in the vicinity of the surface (one or two layers). The effect is thus expected to be reducible to a surface contribution, as observed in the simulations.

(iii) This surface contribution to the free energy is not expected to follow the Gibbs equation due to the presence of the solid layer, explaining the failure of the thermodynamic route to calculate $\gamma$.

(iv) The ordering of the first layer is expected to occur as a sharp transition while increasing the intensity of the fluid-solid interaction, possibly rounded by the finite lateral extension of the nanoplatelet, its surface roughness or the presence of defects. This is compatible with the results of Figure 8, showing that the measured pore-load modulus equals to that given by the thermodynamic model as long as the fluid remains entirely liquid for $\varepsilon_{f\text{-}s} \leq 0.25\ \varepsilon_{f\text{-}f}$, while it is significantly affected above $\varepsilon_{f\text{-}s} \geq 0.5\ \varepsilon_{f\text{-}f}$ where the fluid solidifies at the surface.

## 4. Conclusion

This work presents a molecular simulation examination of the fundamentals of the pore-load measurement technique. This method is widely used to have access to the mechanical properties of nanosized solids, for which direct measurements are otherwise difficult. In order to disentangle geometrical effects (including pore geometry as well as finite wall thickness) from surface contributions, we have performed atomistic simulations that allow comparing quantitatively the observed solid deformation with that expected from the mechanical properties of the elastic solid which can be calculated independently. The model is chosen simple but realistic. The nanoplatelet is shaped out of an fcc solid with elastic constants close to that of the bulk silicon, and the fluid is van der Waals like with parameters corresponding to heptane, used in experiments. The fluid-solid interaction is varied between 0.1 and 1.0 times the fluid-fluid interaction, ranging from "hydrophobic-like" to "hydrophilic-like" surfaces. The temperature is above the triple point, so that the fluid is in the stable liquid phase (bulk).

The molecular simulations show that the pore-load modulus deduced from the deformation of the solid immersed in the liquid at a given pressure or chemical potential may be significantly larger than the value expected from the mechanical properties of the dry nanoplatelet. It is shown however that the expected behavior is recovered if the fluid-wall interaction is attenuated. Analysis of the influence of the nanoplatelet thickness has shown that the discrepancy between the observed pore-load modulus and the expected result reduces to a surface term, which however does not match with the surface free energy from fluid adsorption at the interface as obtained via thermodynamic integration of the Gibbs equation. Atomic structure analysis shows that the fluid exhibits a highly ordered first layer at the solid surface, which disappears when the interaction is reduced. This surface ordering is likely to be the source of the unexpected deformation of the nanoplatelet, since it is a surface effect that is expected to affect the mechanical response of the system, and it is not expected to follow the Gibbs equation.

This ordering of the first layer is expected to be relevant for real systems,[85] in particular those with good affinity of the fluid for the solid and atomically smooth surfaces. Note that the presence of an atomically structured wall may influence the phenomenon,[82,86] but is not required,[84] and is thus expected to be relevant for large molecules.[83] Considering the fact that only the first adsorbed layer will affect the induced deformation, the effect is expected to be stronger for the smallest nanopores.

## Acknowledgment

The author acknowledges fruitful discussions with A. Grosman and E. Rolley on experimental and theoretical issues regarding adsorption-induced deformation, which motivated this study.

**TABLE 1:** Lennard-Jones parameters for the solid-solid (s-s), fluid-fluid (f-f) and fluid-solid (f-s) interaction potentials (eq 1).

|  | s-s | f-f | f-s |
|---|---|---|---|
| $\varepsilon$ (kJ/mol) | 73.2 | 4.20 | 0.42-4.20 |
| $\sigma$ (nm) | 0.3518 | 0.600 | 0.600 |
| cutoff (nm) | 1.4072 | 1.800 | 1.800 |

**TABLE 2:** Numerical values of the compliance matrix elements for the nanoplatelet (see eq 4-6), determined by uniaxial mechanical tests along the x-direction (see section 3.2). For comparison, the corresponding quantities for the bulk solid are also given.

|  | $E_x = \frac{\sigma_{xx}}{\varepsilon_{xx}}$ (GPa) | $\nu_{yx} = \nu_{xy} = -\frac{\varepsilon_{yy}}{\varepsilon_{xx}}$ | $\frac{E_z}{\nu_{zx}} = \frac{E_x}{\nu_{xz}} = -\frac{\sigma_{xx}}{\varepsilon_{zz}}$ (GPa) | $\nu_{xz}$ |
|---|---|---|---|---|
| **nanoplatelet** | 162 | 0.457 | 531 | 0.305 |
| **bulk solid** | 165 | 0.360 | 458 | 0.360 |

**TABLE 3:** Liquid pressure $P_\text{L}$ of the bulk fluid and nanoplatelet deformation given by the molecular simulations as a function of the logarithm of the imposed activity z.

| ln(z) | **-11.175** | **-10.83** | **-10.485** | **-10.200** | **-9.930** | **-9.375** |
|---|---|---|---|---|---|---|
| $P_\text{L}$ **(bar)** | -122 | -63 | -2.5 | 49 | 98 | 202 |
| $\varepsilon_{xx} = \varepsilon_{yy}$ | $-5.65\times10^{-5}$ | $-1.85\times10^{-5}$ | $8.86\times10^{-6}$ | $2.80\times10^{-5}$ | $4.99\times10^{-5}$ | $8.91\times10^{-5}$ |

Figure 1

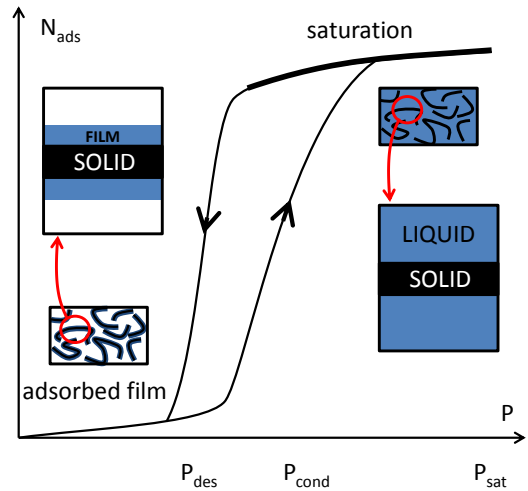

**Figure 1.** Typical adsorption-desorption isotherm for a nanoporous material, giving the amount of fluid adsorbed as a function of the pressure of the gas outside the material (P < $P_{sat}$, the saturated vapor pressure). Three regions can be distinguished. The lower reversible branch (at low pressure) corresponds to a film adsorbed at the walls. The intermediate region exhibits adsorption-desorption hysteresis, the emptying occurring at a pressure lower than condensation (arrows). The upper reversible branch (thick line) corresponds to liquid saturating the nanopores. The pore-load modulus is measured along that branch. The insets show magnifications around a solid wall.

Figure 2

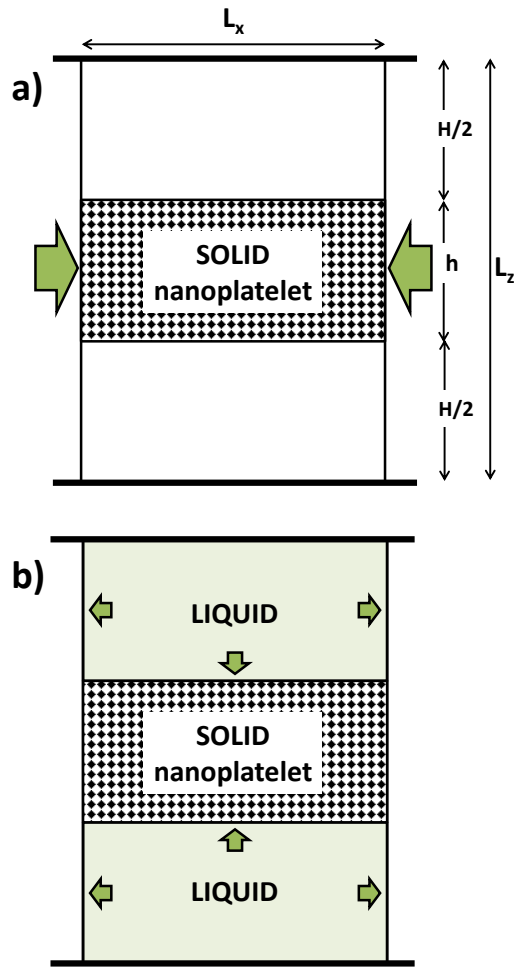

**Figure 2.** General presentation of the simulation box. $L_z$ = 10 nm is constant (thick lines), while $L_x$ and $L_y$ are allowed to fluctuate (isobaric ensemble; thin lines). The nanoplatelet thickness is h = 3.26 nm. a) an external pressure (arrows) is imposed to measure the elastic constants of the nanoplatelet. b) a liquid at imposed chemical potential is introduced; the external pressure is set to zero; the arrows materialize the internal pressure of the liquid.

Figure 3

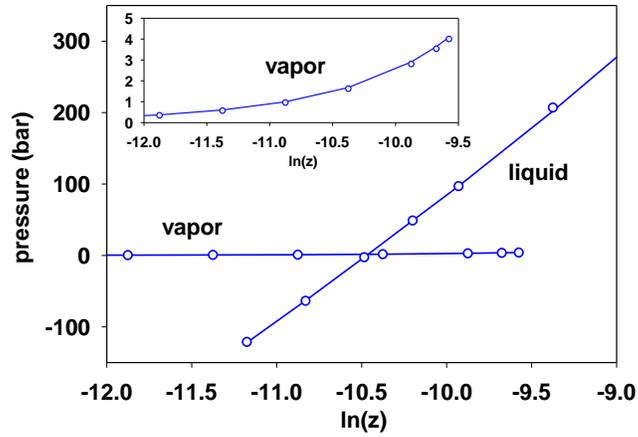

**Figure 3.** Symbols: Grand Canonical Monte Carlo results for the pressure of the bulk liquid and vapor phases of the Lennard-Jones fluid (sections 2.3.2 and 3.1), as a function of the logarithm of the activity (see text). Lines are guide to the eye. Inset: enlargement of the vapor branch. The crossover point corresponds to the coexistence between the vapor and liquid phases ($ln(z) \approx -10.46$; $P_{\text{sat}} \approx 1.6$ bar).

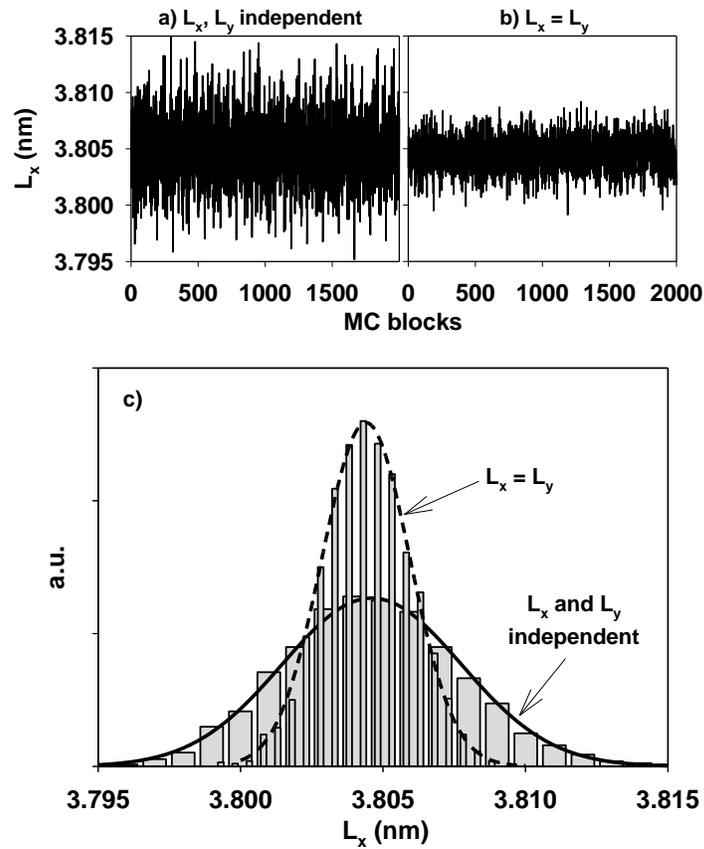

**Figure 4.** a) Fluctuations in the free (no fluid and zero external pressure) nanoplatelet dimension $L_x$ during a Monte Carlo simulation run (2000 blocks of $10^6$ MC steps). b) The geometrical constraint $L_x = L_y$ is imposed. c) Histograms of the fluctuations of $L_x$ corresponding to the situations a and b, given by the simulations (vertical bars) and their Gaussian fits (lines).

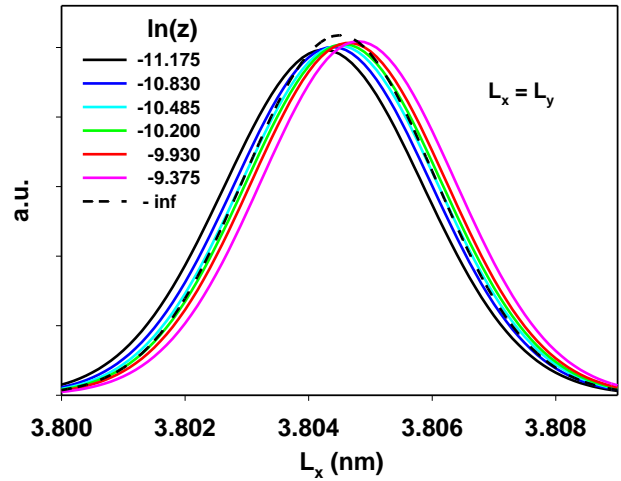

**Figure 5.** Solid lines: Gaussian fits of the fluctuations of the dimension $L_x = L_y$ during a pore-load measurement at various liquid activity z. Dashed line: fluctuations of the dry nanoplate (see Figure 4c).

Figure 6

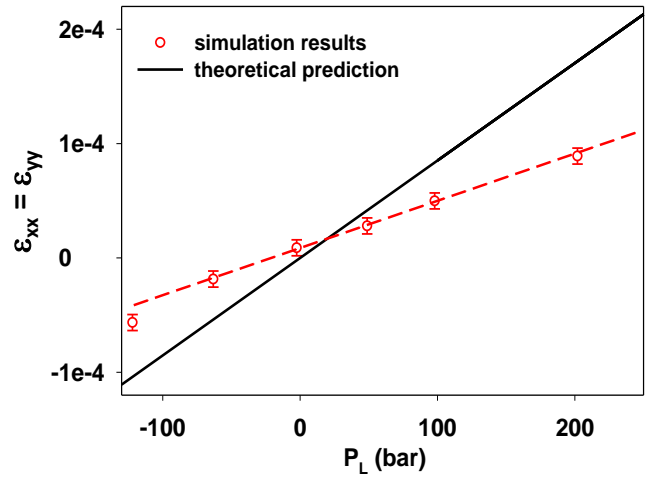

**Figure 6.** Symbols: Monte Carlo simulation results of the nanoplatelet deformation as a function of the fluid pressure in the liquid phase (see Table 3; the dashed line is a guide to the eye). Solid line: theoretical prediction based on the thermodynamic approach (eq 5).

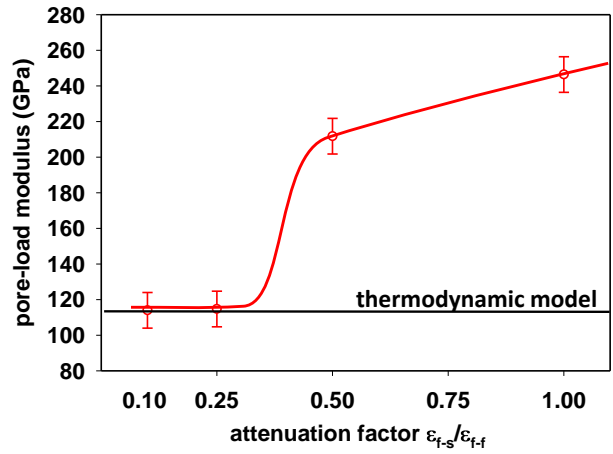

**Figure 7.** Symbols: pore-load modulus obtained by numerical simulations as a function of the attenuation factor applied to the intensity of the fluid-solid interaction $\varepsilon_{f\text{-}s}$ (relative to fluid-fluid interaction $\varepsilon_{f\text{-}f}$). The smooth solid line is a guide to the eye. The horizontal straight line is the result (113 GPa) given by the thermodynamic model (eq 6) and the mechanical parameters of the nanoplatelet (Table 2).

Figure 8

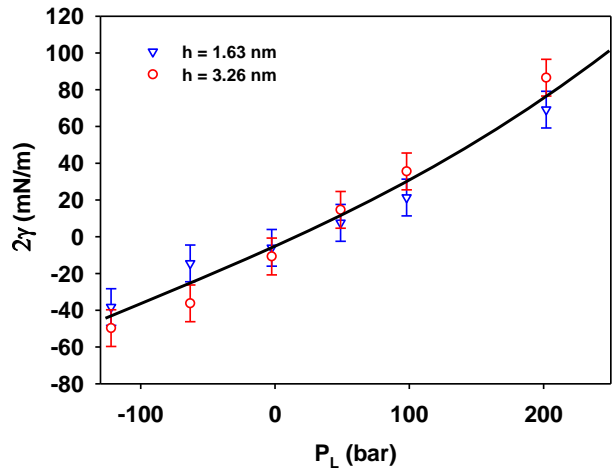

**Figure 8.** Symbols: surface tension $\gamma$ deduced from the measured deformations obtained by molecular simulations (eq 5) for two values of the nanoplate thickness $h$= 3.26 nm (6 unit cells, circles), and $h$= 1.63 nm (3 unit cells, triangles). The factor 2 takes into account the two faces of the nanoplatelet.

Figure 9

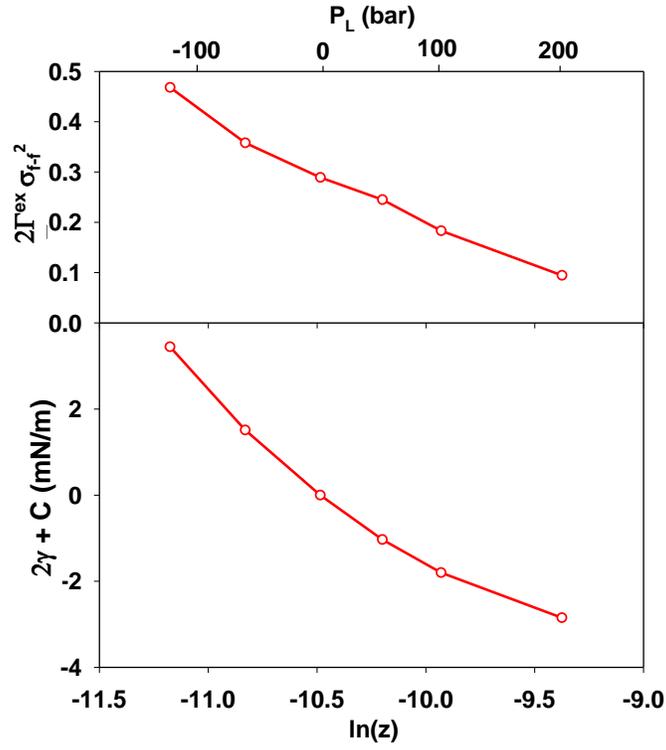

**Figure 9.** Reduced excess adsorbed fluid density $\Gamma^{ex}\sigma_{f\text{-}f}^2$ (upper panel) and the corresponding excess free energy from eq 8 (lower panel), as a function of $\ln(z)$ (lower scale) or pressure (upper scale). The factors 2 take into account the two faces of the nanoplatelet.

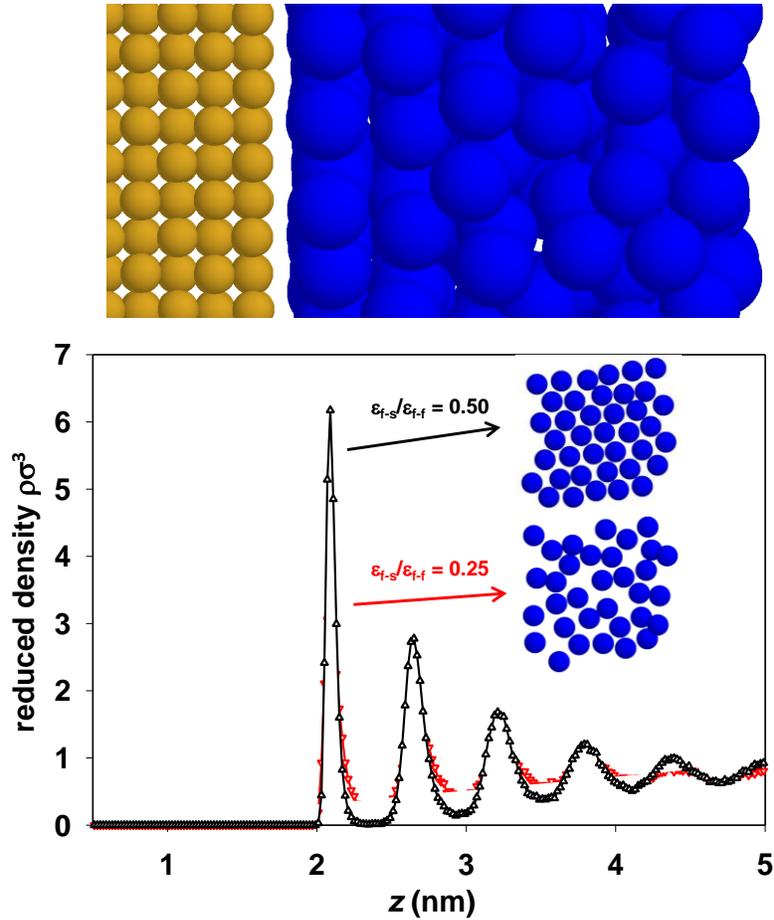

**Figure 10.** Upper panel: molecular configuration showing the liquid (large spheres) in the vicinity of the solid wall (small spheres). Lower panel: local density profile averaged in x and y directions, as a function of z, for two interaction intensities: $\varepsilon_{f\text{-}s}/\varepsilon_{f\text{-}f} = 0.50$ (up triangles), and $\varepsilon_{f\text{-}s}/\varepsilon_{f\text{-}f} = 0.25$ (down triangles). The insets display the first fluid layer in contact with the solid wall.